\documentclass{jpp}
\usepackage{graphicx}
\usepackage[dvipsname]{xcolor}
\usepackage{amsmath,amsfonts}
\usepackage{subfigure}
\usepackage{bm}
\usepackage[outdir=./]{epstopdf}
\usepackage[only,llbracket,rrbracket]{stmaryrd}
\usepackage[colorlinks=true, linkcolor=olive,citecolor=olive]{hyperref}
\usepackage{upgreek}

\shorttitle{Plasma potential shaping in the Large Plasma Device}
\shortauthor{ R. Gueroult, S. K. P. Tripathi, F. Gaboriau, T. R. Look and N. J. Fisch}

\title{Plasma potential shaping using end-electrodes in the Large Plasma Device}

\author{R. Gueroult\aff{1}\corresp{\email{renaud.gueroult@laplace.univ-tlse.fr}}, S. K. P. Tripathi\aff{2}, F. Gaboriau \aff{1}, T. R. Look\aff{2}, 
  \and N. J. Fisch\aff{3}}

\affiliation{\aff{1}LAPLACE, Universit\'{e} de Toulouse, CNRS, INPT, UPS, 31062 Toulouse, France \aff{2}Department of Physics and Astronomy, University of California, Los Angeles, Los Angeles, California 90095, USA
\aff{3}Department of Astrophysical Sciences, Princeton University, Princeton NJ 08540, USA}

\definecolor{tabblue}{HTML}{1f77b4}
\definecolor{taborange}{HTML}{ff7f0e}
\definecolor{tabgreen}{HTML}{2ca02c}
\definecolor{tabred}{HTML}{d62728}
\definecolor{tabpurple}{HTML}{9467bd}
\definecolor{tabbrown}{HTML}{8c564b}
\definecolor{tabpink}{HTML}{e377c2}
\definecolor{tabgray}{HTML}{7f7f7f}
\definecolor{tabolive}{HTML}{bcbd22}
\definecolor{tabcyan}{HTML}{17becf}

\begin{document}

\maketitle

\begin{abstract}
We perform experiments in the Large Plasma Device (LAPD) at the University of California, Los Angeles, studying how different end-electrode biasing schemes modify the radial potential profile in the machine. We impose biasing profiles of different polarities and gradient signs on a set of five concentric electrodes placed $12$~m downstream from the plasma source. We find that imposing concave-down profiles (negative potential radial gradient) on the electrodes create radial potential profiles halfway up the plasma column that are comparable to those imposed on the electrodes, regardless of the biasing polarity. On the other hand, imposing concave-up profiles (positive potential radial gradient) leads to non-monotonic radial potential profiles. This observation can be explained by the current drawn through the electrodes and the parallel plasma resistivity, highlighting their important role in controlling the rotation of plasma. Concave-down plasma potential profiles, obtained by drawing electrons on the axis, are predicted to drive azimuthal drift velocities that can approach significant fractions of the ion sound speed in the central region of the plasma column.
\end{abstract}

\section{Introduction}

Producing and controlling rotation in plasmas is attractive for a number of applications~\citep{Lehnert1971}. For example, plasma rotation could prove essential to advancing magnetic confinement fusion in both linear~\citep{Ellis2001,Ellis2005,RomeroTalamas2021,Endrizzi2023} and toroidal~\citep{Rax2017} geometries. Plasma rotation control is also a key element in advancing plasma mass separation technologies~\citep{Zweben2018,Gueroult2018}, both in plasma centrifuges for isotope separation~\citep{Krishnan1981,Fetterman2009,Borisevich2020} and cross-field rotating collisionless separator concepts targeting new separation needs~\citep{Ohkawa2002,Fetterman2011,Gueroult2014,Liziakin2022}. Finally, controlling plasma rotation in laboratory experiments provides unique opportunities for investigating the role of plasma rotation in affecting the dynamics of laboratory and space plasmas --- such as the formation of accretion disks \citep{Flanagan2020,ValenzuelaVillaseca2023} and propagation characteristics of plasma waves~\citep{Gueroult2019a,Gueroult2023}.

In magnetized plasmas, a possible means of controlling rotation is through cross-field drift driven by a perpendicular electric field. While new promising schemes harnessing beams and waves have been proposed to establish this perpendicular electric field~\citep{Putvinskii1981,Rax2023}, a long-studied solution is to use biased electrodes. One option is to bias electrodes positioned at the outer edge of the plasma, referred to as limiters, to impose a radial potential difference. Although this scheme has been shown to effectively establish an electric field and rotation in the plasma edge~\citep{Schaffner2012}, the electric field and rotation in the bulk plasma is due to plasma self-reorganisation~\citep{Weynants1993} and cannot be directly controlled.
Another option, which has been surmised to provide greater control over the electric field in the plasma bulk, is the end-electrodes scheme originally proposed by~\cite{Lehnert1970,Lehnert1973}. The principle is to use independently biased electrodes, on which the plasma column terminates, to finely shape the perpendicular electric field in the plasma bulk. Despite having been implemented in numerous experimental setups over the years (see \emph{e.~g.} \cite{Gueroult2019} and references therein), the effectiveness of end-electrode biasing in controlling the perpendicular electric field across plasma regimes remains unclear.
Two outstanding issues are determining the fraction of the imposed bias lost in the sheath and how much the potential varies along magnetic field lines~\citep{Trotabas2022}. In this study, we report the results of an experimental campaign using end-electrodes carried out in the Large Plasma Device (LAPD) at the University of California, Los Angeles in August 2023 that shed light on this problem.

As mentioned above, numerous experiments have used biased end-electrodes. However, the operating conditions and size of LAPD allow for exploring different regimes. Indeed, it has been shown~\citep{Poulos2019,Gueroult2019b} that an important dimensionless parameter affecting the uniformity of the plasma potential along a field line is
\begin{equation}
\tau = \frac{L}{a}\sqrt{\frac{\sigma_{\perp}}{\sigma_{\parallel}}}
\label{Eq:tau}
\end{equation}
with $L$ and $a$ the plasma column length and radius, respectively, and $\sigma_{\perp}$ and $\sigma_{\parallel}$ the perpendicular and parallel plasma conductivities. Many of the recent end-electrodes biasing experiments used helicon plasmas in mid-size experiments where typically $\tau\geq1$~\citep{Gilmore2011,Gilmore2014,Zhang2015,Gueroult2016a,Liziakin2019,Desangles2021}. However, this regime has been inferred to prohibit effective electrical connection between symmetrical electrodes positioned at opposite ends of the machine~\citep{Gueroult2019b}, contrary to the operating principle envisioned by~\cite{Lehnert1970,Lehnert1973}.  On the other hand the low operating pressure in LAPD typically leads to regimes where $\tau\ll1$, creating opportunities to test these models. 

This paper is organised as follows. In Section~\ref{Sec:Expe} we begin by introducing the experimental setup used in this experimental campaign, and how it both relates to and differs from past experiments. In Sec.~\ref{Sec:Baseline} we analyse plasma parameter profiles obtained for the selected operating condition prior to inserting the electrodes in the machine, which will serve as a baseline. In Section~\ref{Sec:Biasing} we present results obtained for a set of six complementary biasing scenarios, which are shown to exhibit distinctly different plasma potential profiles. In Section~\ref{Sec:Model} we then propose a simple analysis to explain trends in plasma potential modifications, considering both the voltage drop along field lines and the dynamics of the anode-plasma system, and discuss this model in light of previous contributions. Finally, in Sec.~\ref{Sec:Rotation} we use the plasma potential profiles measured in the different biasing scenarios to infer the corresponding plasma rotation expected from cross-field drift and show how imposing a concave-down biasing profile leads to azimuthal drift velocities approaching a significant fraction of the ion sound speed. These results are briefly discussed in regard to previous studies demonstrating biasing driven rotation in LAPD. Lastly, in Section~\ref{Sec:Conclusion}, we summarise the main findings of this study.

\section{Experimental setup}
\label{Sec:Expe}

In this section, we introduce the main characteristics of the experimental setup used in this campaign, which will be helpful for our upcoming analysis and discussion of biasing results.

\subsection{Biasing setup}

Experiments are carried out in the Large Plasma Device (LAPD) at the University of California, Los Angeles~\citep{Gekelman2016}. LAPD uses a large $38$~cm diameter hot LaB$_{6}$ cathode~\citep{Qian2023} to produce a highly reproducible $20$~m long, $75$~cm diameter magnetized plasma column with a repetition rate of $0.3-1$~Hz. 

\begin{figure}
\begin{center}
\subfigure[Multi-disk electrode]{\includegraphics[height=6cm]{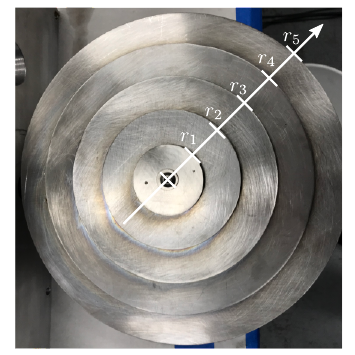}\label{Fig:electrodes_pic}}\hfill\subfigure[Biasing circuit]{\includegraphics[width=6cm]{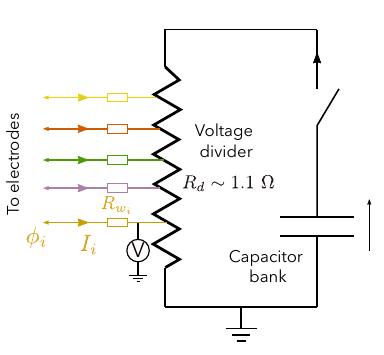}\label{Fig:electrodes_circuit}}
\caption{Picture of the multi-disk electrode used in this experimental campaign \subref{Fig:electrodes_pic} and electric circuit used to bias each electrode \subref{Fig:electrodes_circuit}. The outer radius of each electrode is $r_{i}=2.54 i$~cm, $i\in\llbracket1,5\rrbracket$. $R_{w_{i}}$ is the resistance of the cable between the point of measure near the voltage divider and the disk $\mathcal{E}_{i}$ in the machine. }
\label{Fig:Electrodes}
\end{center}
\end{figure}

\begin{figure}
\begin{center}
\includegraphics[]{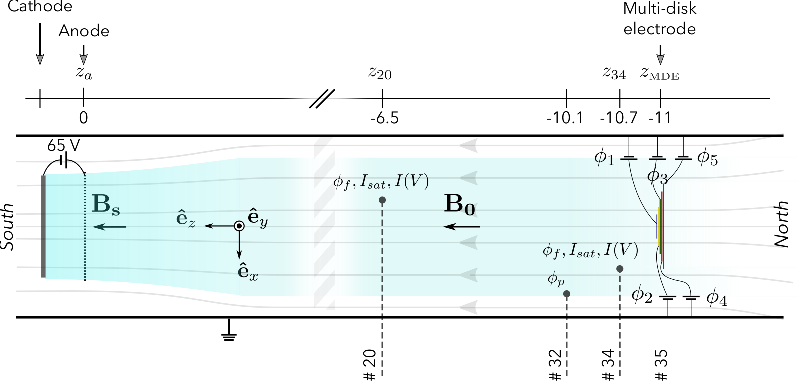}
\caption{Experimental setup used in this campaign. The plasma is created by applying a voltage between the cathode and a mesh anode. The five concentric disk electrodes are installed on port \#35 and biased with respect to the grounded vacuum chamber, with bias $\phi_{i}$ for $i\in\llbracket1,5\rrbracket$. The magnetic field in the source $B_{s}$ is $0.2$~T whereas the field in the main chamber $B_{0}$ is $0.1~$T. }
\label{Fig:Setup}
\end{center}
\end{figure}

In this campaign, we use a set of five stainless steel disks, electrically separated by ceramic spacers, to form a stack of concentric electrodes as shown in figure~\ref{Fig:electrodes_pic}. Each disk, $\mathcal{E}_{i}$ with $i\in\llbracket1,5\rrbracket$, has a radius $r_{i}=2.54 i$~cm. This electrode stack, or multi-disk electrode, is centered radially at the machine port $\#35$ and rotated perpendicular to the magnetic field. As illustrated in figure~\ref{Fig:Setup}, the electrode stack is about $11$~m from the anode. Although this multi-disk electrode setup is similar in design and comparable in size to one previously used in LAPD by \cite{Koepke2008,Koepke2016}, an important difference concerns the biasing scheme on the electrodes. In this study, we bias each electrode with respect to the machine ground, whereas electrodes were biased with respect to each other (or a group with respect to another group) in earlier studies. As highlighted in figure~\ref{Fig:electrodes_circuit}, the voltage $\phi_{i}$ of each electrode $\mathcal{E}_{i}$ is set by positioning pins across a resistive voltage divider ($R_{d}\sim1.1~\Upomega$), which is connected to a capacitor bank through a transistor switch. We note that this experimental setup also shares commonalities with the work by~\cite{Jin2019}, though in this case, the emissive cathode positioned in the machine was biased with respect to the anode.

We also note an important difference in this setup compared to the canonical end-electrodes configuration proposed by \cite{Lehnert1970}. As shown in figure~\ref{Fig:Setup}, one end of the machine is blocked by the anode mesh. As a result, the configuration is not symmetrical. Although it makes it impossible to test the assumption of effective electrical connection between symmetrical end-electrodes for $\tau\ll1$, it is still possible to study how the potential distributes itself along and across field lines in these conditions. In this regard, our study builds on the study by~\cite{Jin2019} by exploring different biasing configurations using multiple electrodes $\phi_{i}$ and both positive and negative biases. 

\subsection{Shot parameters}

\begin{figure}
\begin{center}
\includegraphics[]{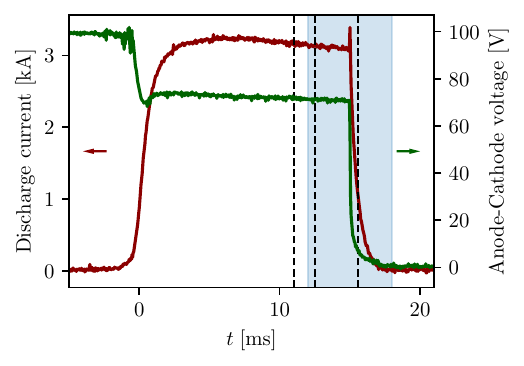}
\caption{Time evolution of the anode-cathode voltage (green, right axis) and discharge current (red, left axis) during a typical shot. The $t=0$ reference corresponds to $I_{d}=1$~kA. The blue region $12\leq t\leq 18$~ms corresponds to the time the electrode bias is on. The vertical dashed lines indicate $3$ different instants during the shot that we will focus on in our data analysis. The first one ($11.1$~ms) is before bias has been turned on, the second one ($12.6$~ms) is during the main discharge with bias on, and the last one ($15.6$~ms) is in the afterglow with bias on. }
\label{Fig:Current_trace}
\end{center}
\end{figure}

With the LAPD cathode upgrade~\citep{Qian2023}, the discharge is fueled by gas puffed near the new LaB$_6$ source. In this campaign, we use helium gas puff with a maximum chamber pressure reaching $3.45\times10^{-5}$~ Torr and a $15~$ms long pulsed discharge with a repetition rate of $1$~Hz. The source magnetic field and magnetic field in the main chamber are respectively $B_{s}=0.2$~T and $B_{0}=0.1$~T, for a field ratio $\rho_{B} \doteq B_{s}/B_{0}=2$. In these conditions, as can be seen in figure~\ref{Fig:Current_trace}, the peak discharge current is $I_{d}\sim3.2$~kA with an anode-cathode voltage difference of about $65$~V. We will verify that it corresponds to a plasma density of about $6\times10^{18}$~m$^{-3}$, which was used as our target operating point during the one-week-long campaign. 

The active phase during which electrodes are biased starts at $12~$ms and ends at $18~$ms (blue region in figure~\ref{Fig:Current_trace}). The active bias phase thus covers both the end of the main discharge during which electrons are still injected in the plasma and the afterglow regime which refers to the phase after the discharge voltage has been switched off~\citep{Gekelman2016}. Since the electron temperature $T_{e}$ drops rapidly in the afterglow, this allows probing the effect of biasing in different plasma conditions.

\subsection{Diagnostics}

Plasma parameters in the machine are inferred through a suite of probe diagnostics. In this campaign, four-tip probes are mounted on probe drives on ports $\#20$ and $\#34$, which correspond to distances of about $5.5~$m and $30$~cm from the multi-disk electrode. The tip area is about $4$~mm$^{2}$. On each of these probes, three tips are used. Two of these three tips are used to acquire respectively the ion saturation current $I_{sat}$ and the floating potential $\phi_{f}$. The last tip is used as a swept Langmuir probe to acquire $I(V)$ characteristics. Plasma potential estimates $\phi_{p}$ are obtained from $I(V)$ probe characteristics using spline fitting and standard inflection-point method~\citep{Godyak2015}. The electron temperature is determined from a linear fit of the logarithm of the electron current as a function of bias for biases between the floating potential and the inflection point. Practically, $200$~ms long $I(V)$ Langmuir sweeps separated by a $100$~ms pause were used, for a total of $23$ sweeps starting at $t=11.1~$ms. In addition, an emissive probe~\citep{Martin2015} is mounted on a third probe drive on port $\#32$, which is about $90$~cm away from the multi-disk electrode. The heating current is first adjusted to ensure symmetrical $I(V)$ curves, whereafter the plasma potential $\phi_{p}$ is taken to be the floating potential of the emissive electrode.  All three probe drives are computer-controlled, making it possible to obtain 2D maps of plasma parameters at three axial locations ($\#20$, $32$ and $34$). In practice we will mostly focus on results from planes at $\#20$ and $34$, and will verify that the plasma potential estimated from the emissive probe on port $\#32$ is consistent with estimates obtained from $I(V)$ data on port $\#34$. Absolute plasma density estimates are simply deduced from $I_{sat}$ data after calibration using microwave interferometric data taken on port $\#20$. Variations in $I_{sat}$ are therefore entirely attributed to variations in density $n$, while they could also result from variations in electron temperature $T_{e}$. Although crude, this assumption is supported by the weaker dependence of the ion saturation on the temperature ($\propto\sqrt{T_{e}}$) compared to the density ($\propto n_{e}$), and the relatively limited variations in $T_{e}$ which will be observed. Lastly, the data presented here is averaged over five shots taken at the same spatial position, with error bars quantifying the standard deviation over these five shots.

Besides plasma parameters and general machine parameters, the voltage $\phi_{i}$ of each electrode $\mathcal{E}_{i}$ with $i\in\llbracket1,5\rrbracket$,  the current $I_{i}$ going through each of these electrodes, as well as the anode potential $\phi_{a}$, are recorded throughout the shot. For practical reasons, the voltages are measured right next to the voltage divider illustrated in figure~\ref{Fig:electrodes_circuit}, which is positioned some meters away from the electrodes, outside the chamber. Due to the finite resistance of these long connectors, the actual electrode potential $\phi_{i}$ thus has to be corrected given the current $I_{i}$ and the resistance $R_{w_i}$ of each of these connectors. These resistors were measured before the campaign with the multi-disk electrode outside the chamber, with $R_{w_i} = 0.160$, $0.157$, $0.167$, $0.162$ and $0.148~\Upomega$ for $i\in\llbracket1,5\rrbracket$.

\section{Baseline parameters}
\label{Sec:Baseline}

\subsection{Baseline profiles}

\begin{figure}
\begin{center}
\includegraphics[width=13.5cm]{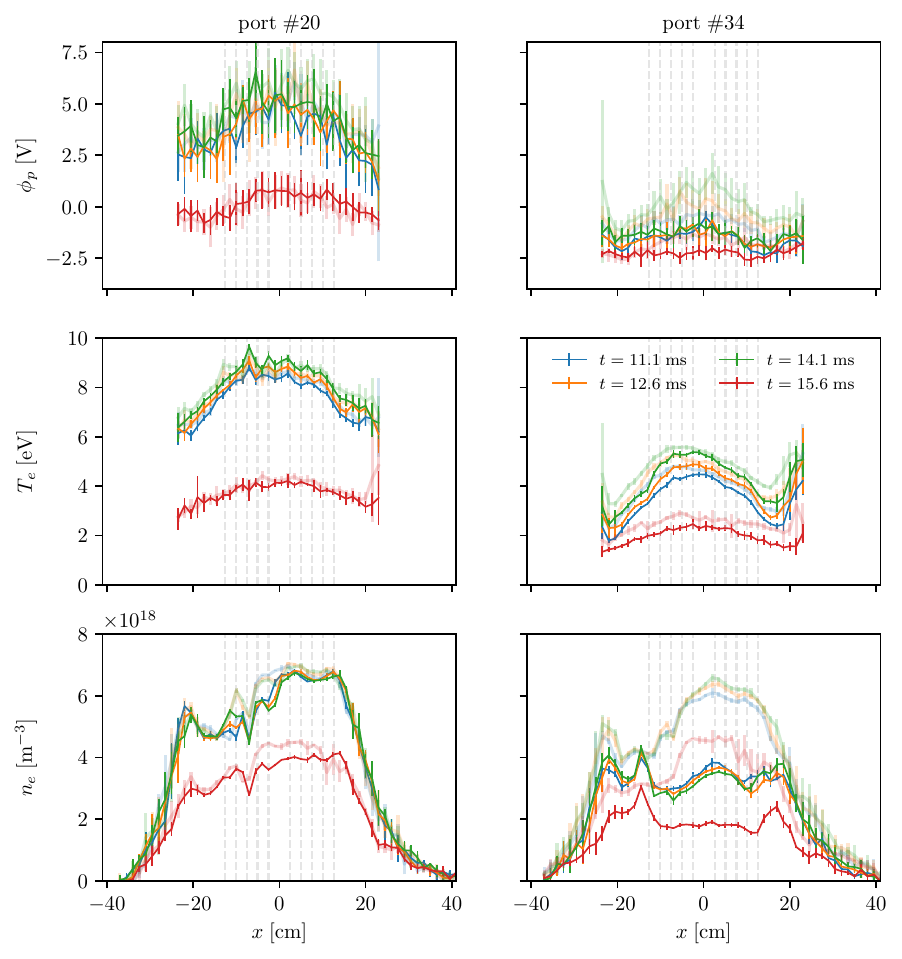}
\caption{Radial profiles of the plasma potential (top panels) and the electron temperature (middle panels) inferred from Langmuir probe sweeps, and of the floating potential (top, cross) and of the density (bottom) deduced from floating probe and saturation current (without $T_{e}$ corrections) at four times during the discharge, and on two ports: port $\#20$ (left) and port $\#34$ (right). The lighter curves correspond to the data taken with the electrodes outside the machine. The vertical dashed lines represent the electrode's position (when inserted). }
\label{Fig:Baseline_no_electrode}
\end{center}
\end{figure}

With our goal to highlight the effect of biasing on plasma parameters---particularly on the plasma potential---we briefly discuss the plasma parameter profiles measured for the selected operating point, first with the multi-disk electrode pulled out of the chamber and then with the multi-disk electrode in the machine but disconnected. This dataset is shown in figure~\ref{Fig:Baseline_no_electrode}. 

Starting with density (lower panel in figure~\ref{Fig:Baseline_no_electrode}), we measure a maximum of about $6\times10^{18}$~m$^{-3}$ in the center during the main discharge, and a decrease with time after the discharge has been turned off ($t=15~$ms). Both in the main discharge and in the afterglow, the density drops rapidly for $r\geq25$~cm. This is consistent with a plasma column radius equal to the cathode radius. Indeed, the cathode radius is $r_{c}=19~$cm, which considering flux conservation and the magnetic field ratio $\rho_{B}\doteq B_{s}/B_{0}=2$ between the source and the main chamber gives a plasma column radius $r_{c}\sqrt{\rho_{B}}\sim27$~cm. The density data also shows an asymmetry, with density for positive $x$ larger than density for negative $x$. This asymmetry does not vanish when averaging over time and is also larger than the shot-to-shot deviation, as shown by the error bars. Although the origin of this asymmetry remains to be determined, it might be evidence for probe perturbation. Negative $x$ indeed corresponds to the far side of the machine, for which the probe arm extends past the central point. This could lead to perturbations, in particular if the plasma is rotating.

Moving on to the middle panel in figure~\ref{Fig:Baseline_no_electrode}, the electron temperature deduced from $I$-$V$ characteristics is about $8$~eV in the core during the main discharge on port $\#20$, and drops to about $4$~eV at the same location after the discharge has been turned off. Closer to the electrodes on the port $\#34$, a similar radial profile is observed but with an overall axial drop of about $5$~eV in the main discharge and $2$~eV in the afterglow, leading to peak temperatures of respectively $4$ and $2$~eV. 

Lastly, the plasma potential $\phi_{p}$ (top panel in figure~\ref{Fig:Baseline_no_electrode}) is relatively flat across the plasma column. It is respectively positive and negative by a few volts on port $\#20$ and on port $\#34$. The axial voltage drop between these two ports is about $5$~V in the main discharge and about $2$~V in the afterglow. This axial variation is consistent with the voltage drop expected from the axial temperature variation highlighted just above. Indeed the electron momentum equation gives
\begin{equation}
en_{e}\frac{\partial \phi_{p}}{\partial z} - \frac{\partial (n_{e}T_{e})}{\partial z} +en_{e}\eta_{{\parallel}}j_{z}= 0
\label{Eq:elec_mom}
\end{equation}
so that neglecting axial current
\begin{equation}
\frac{\partial \phi_{p}}{\partial z}  =  \frac{1}{e}\frac{\partial T_{e}}{\partial z}+\frac{T_{e}}{en_{e}}\frac{\partial n_{e}}{\partial z}>0.
\end{equation}
Over distances short compared to the axial density gradient length, which we verify in the lower panel in figure~\ref{Fig:Baseline_no_electrode} corresponds to the distance between port $\#20$ and port $\#34$, the first term on the right hand side dominates. The axial voltage drop is then equal to lowest order to the axial temperature variation.

Finally, we note that introducing the disconnected electrodes to the plasma produces a noticeable effect at the port $\#34$ which is located closest to the electrodes. A significant density depletion, consistent with enhanced losses, occurs in the central region where the field lines connect to the electrodes. A closer examination here also reveals a modest drop in $T_{e}$---and consequently $\phi_{p}$. 

\subsection{Currents}

From the plasma density $n_{e}$ and electron temperature $T_{e}$ measured above, we can deduce estimates for the ion saturation current density $j_{is}\doteq en_{e}c_{s}$ with $c_{s}\doteq\sqrt{k_{B}T_{e}/m_{i}}$ the ion sound speed. Using $n_{e}=5\times10^{18}$~m$^{-3}$ and $T_{e}=5$~eV (see table~\ref{Tab:baseline}) as a baseline operating point in this campaign, we obtain a characteristic ion saturation current density 
\begin{equation}
j_{is}^{\diamond}=8.7~\textrm{kA.m}^{-2}.
\label{Eq:j_is_diamond}
\end{equation}

\begin{table}
\begin{center}
\begin{tabular}{c c}
Parameter & Value\\
\hline
Gas & helium\\
Discharge current $I_{d}\quad$[kA] & 3.2 \\
Main chamber magnetic field $B_{0}\quad$ [T] & 0.1\\
Magnetic field ratio $\rho_{b}$ & $2$\\
Plasma density $n_{e}\quad[\textrm{m}^{-3}]$ & $5~10^{18}$ \\
Electron temperature $T_{e}\quad$[eV] & $5$\\
Plasma radius$\quad$[cm] & $25$\\
Plasma length$\quad$[m] & $11$
\end{tabular}
\caption{Typical machine and plasma parameters for the operating point targeted in this campaign. }
\label{Tab:baseline}
\end{center}
\end{table}

For reasons that will become clear later, it is informative to compare this current to the current density deduced from the discharge current $I_{d}$. Considering a uniform emission on the cathode, and accounting for the anode transparency $\iota=0.66$~\citep{Qian2023}, the current density of primary electrons injected in the plasma through the anode during the main discharge is 
\begin{equation}
j_{inj} = -\frac{\iota I_{d}}{\pi\rho_{B}r_{c}^{2}}
\label{Eq:j_inj}
\end{equation} 
where $\rho_{B}$ is the field ratio previously defined. For the discharge current $I_{d}\sim3.2$~kA  shown in figure~\ref{Fig:Current_trace} and $\rho_{B}=2$ one gets $j_{inj}=-9.3~\textrm{kA.m}^{-2}$. 

Although the amplitude of these currents will vary with the location through the spatial dependence of $T_{e}$ and $n_{e}$ on one hand, and with the exact discharge current on the other hand, the important point here is to note that in the conditions investigated here these two terms are of comparable magnitude.

\section{Effect of biasing on the radial profile of plasma parameters}
\label{Sec:Biasing}

\subsection{Biasing profiles}

\begin{figure}
\begin{center}
\includegraphics[width=13.5cm]{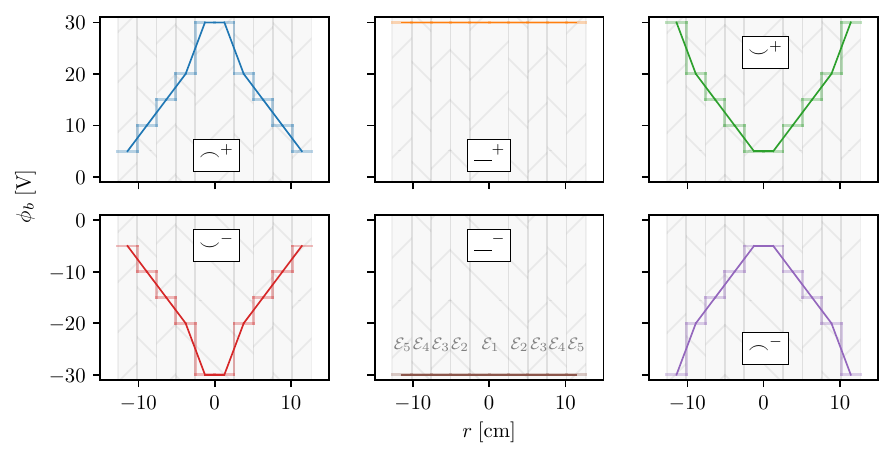}
\caption{Potential profile imposed on the electrodes for the six biasing scenarios tested in this campaign, as listed in table~\ref{Tab:cases}. The thin solid coloured lines represent the continuous targeted profile, while the thicker coloured segments represent the actual steplike profile imposed on the electrodes. The grey hatched regions represent the position of the different electrodes $\mathcal{E}_{i}$.}
\label{Fig:Biasing_scheme}
\end{center}
\end{figure}

We choose six biasing profiles to impose on the multi-disk electrode to study the effect of differential electrode biasing on plasma parameters. Our motivation for choosing these six specific biasing scenarios, which are listed in table~\ref{Tab:cases} and illustrated in figure~\ref{Fig:Biasing_scheme}, is that they cover different polarities with respect to machine ground as well as different directions of the radial gradient of the potential. In other words, we examine the effect of both positive and negative biases, as well as concave-up and -down potential profiles. The amplitude in all cases is $30$~V, though as mentioned above the resistance of the connectors is such that the actual bias on the electrode can be slightly larger or smaller depending on the current drawn by this particular electrode.

\begin{table}
\begin{center}
\begin{tabular}{c c c}
Biasing profile & Shorthand & Color code \\
 \hline
Positive, concave-down (negative gradient) & $\frown^{+}$ & \textcolor{tabblue}{\rule{.4cm}{0.1cm}}\\
Positive, flat & \rule{0.3cm}{0.015cm}$^{+}$ & \textcolor{taborange}{\rule{.4cm}{0.1cm}}\\
Positive, concave-up (positive gradient) & $\smile^{+}$ & \textcolor{tabgreen}{\rule{.4cm}{0.1cm}}\\
Negative, concave-up (positive gradient) & $\smile^{-}$ & \textcolor{tabred}{\rule{.4cm}{0.1cm}}\\
Negative, flat & \rule{0.3cm}{0.015cm}$^{-}$ &\textcolor{tabbrown}{\rule{.4cm}{0.1cm}}\\
Negative, concave-down (negative gradient) & $\frown^{-}$ & \textcolor{tabpurple}{\rule{.4cm}{0.1cm}}\\
\end{tabular}
\caption{Six biasing scenarios studied in this campaign. The amplitude is $30$~V in all cases. }
\label{Tab:cases}
\end{center}
\end{table}

\begin{figure}
\begin{center}
\includegraphics[width=13.5cm]{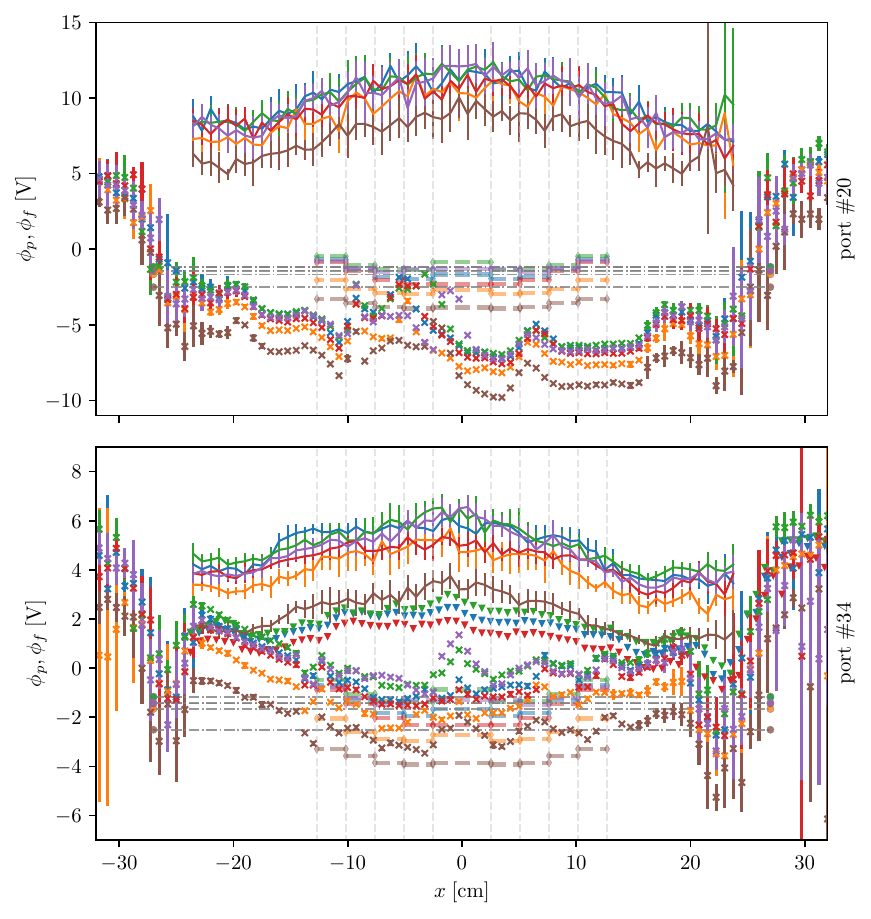}
\caption{Plasma (solid lines) and floating (cross markers) potential profiles on two ports ($\#20$ top, $\#34$ bottom) at $t=11.1$~ms (before biasing) for the six biasing scenarios listed in table~\ref{Tab:cases}. The vertical dashed lines represent the electrodes position. The thicker and lighter colored horizontal lines represent the electrodes potential $\phi_{i}$ at that time. The horizontal grey dashed lines with coloured circles symbols represent the anode potential $\phi_{a}$. The triangle symbol in the bottom panel corresponds to plasma potential measurement from emissive probe on port $\#33$ (when available). }
\label{Fig:t0_FP_6base}
\end{center}
\end{figure}

A consequence of this resistance, and of the biasing circuit shown in figure~\ref{Fig:electrodes_circuit}, is that the plasma potential profiles already show a weak dependence on the particular biasing scenario implemented before the active biasing phase. This is illustrated in figure~\ref{Fig:t0_FP_6base}, which corresponds to $t=11.1$~ms, that is $0.9$~ms before the biasing power supply is turned on. Indeed, even if the bias is off, electrodes are not floating. They are connected through the connectors and the voltage divider. Because the position of a given pin on the voltage divider differs with the different biasing scenarios, the resistance between a given electrode and ground also differs. This leads to a different electrode bias, and from there to a different plasma potential in front of this electrode. Note that the larger error bars seen in figure~\ref{Fig:t0_FP_6base} for $r\gtrsim 25$~cm corresponds to the region beyond the projected cathode radius $r_{c}\sqrt{\rho_{B}}\sim27$~cm where the density drops rapidly.

We also verify in figure~\ref{Fig:t0_FP_6base} that, while there is a small offset of about $2$~V between the two datasets, the plasma potential estimated from $I(V)$ sweeps on port $\#34$ (solid lines) is consistent with that obtained about $60$~cm away from the emissive probe on port $\#32$ (triangle symbols). The radial dependence predicted by both methods is notably remarkably similar. This is particularly noteworthy in that these two datasets were obtained in different runs (days apart), so that small variations in plasma parameters cannot be excluded.

\subsection{Potential profiles under active biasing}

\begin{figure}
\begin{center}
\includegraphics[width=13.5cm]{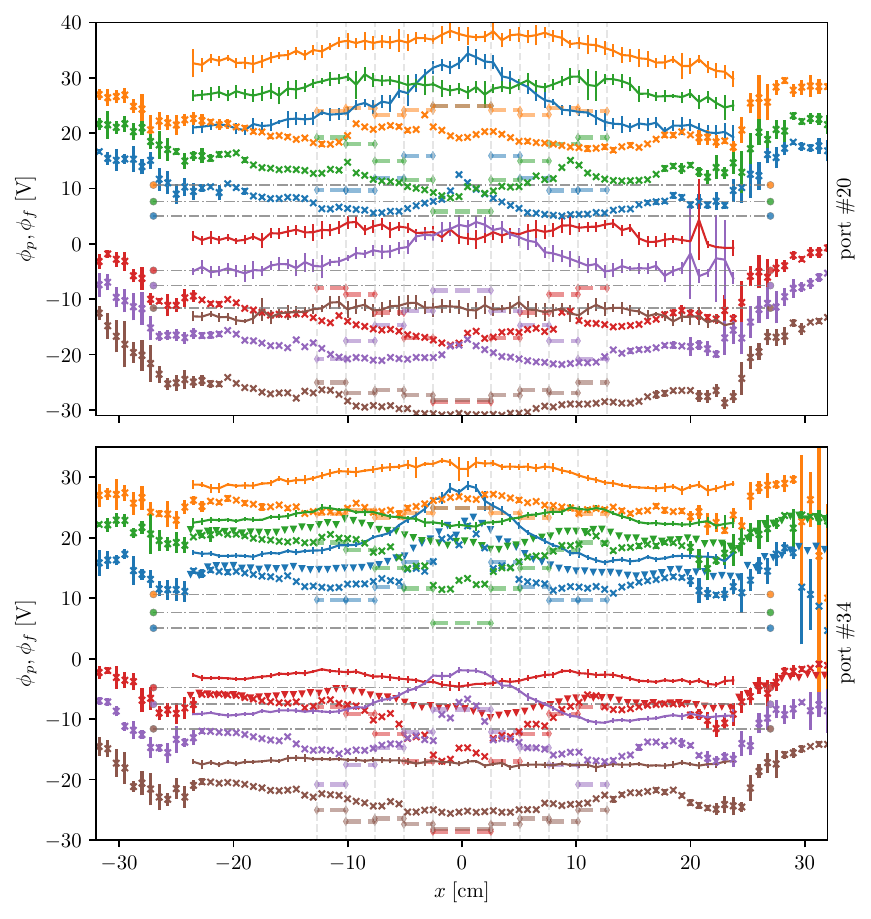}
\caption{Plasma (solid lines) and floating (cross markers) potential profiles on two ports ($\#20$ top, $\#34$ bottom) at $t=12.6$~ms (\emph{i.~e.} in the main discharge during biasing) for the six biasing scenarios listed in table~\ref{Tab:cases}. The vertical dashed lines represent the electrodes position. The thicker and lighter colored horizontal lines represent the electrodes potential $\phi_{i}$ at that time. The horizontal grey dashed lines with coloured circles symbols represent the anode potential $\phi_{a}$. The triangle symbol in the bottom panel correspond to plasma potential measurement from emissive probe on port $\#33$ (when available). }
\label{Fig:t1_FP_6base}
\end{center}
\end{figure}

As shown in figure~\ref{Fig:t1_FP_6base}, biasing has a very noticeable effect on the radial profile of plasma potential during the main discharge ($t=12.6$~ms). We can make a number of general observations before attempting to explain this plasma response in the next section.

First, the amplitude of the plasma potential variations in response to biasing is significant. For the conditions used here, one finds $-15\leq\phi_{p}\leq35$~V with active biasing, compared to $5\leq\phi_{p}\leq10$~V with biasing off. The amplitude of this shift is thus comparable to the maximum bias of $\pm30$~V applied on the electrodes. This suggests that only a modest fraction of the applied bias is lost through the sheaths~\citep{Poulos2019,Trotabas2022}. 
Second, the radial profile of the plasma potential is found to track---if not to be driven by---the applied bias profile.  More precisely, figure~\ref{Fig:t1_FP_6base} shows that it is the floating potential $\phi_{f}(r,z_{34})$ rather than the plasma potential $\phi_{p}(r,z_{34})$ which tracks the applied bias $\phi_{b}(r)$, and that concave-down profiles which have negative radial gradients (blue and purple) appear to be more easily passed into the plasma column than those with positive gradients (red and green). 
Lastly, we note that these variations in plasma potential are accompanied by smaller yet significant variations in anode potential $\phi_{a}$. The anode potential is indeed observed to vary in these conditions from $-11$ to $11$~V depending on the biasing scenario.

Looking more closely at the ordering between the plasma potential and the electrode bias, we find that for constant applied bias across the electrodes (orange and brown) the plasma potential is larger than the bias on any electrode. The difference between the plasma potential on the port $\#34$ and the applied bias on the electrodes is then relatively uniform radially. For non-uniform biasing profiles, the overall profile appears to be set by the condition that the plasma potential is to be slightly more positive than the most positive electrode in the region in front of this electrode. Because the plasma potential does not precisely follow the applied bias, the voltage drop between a given electrode $\mathcal{E}_{i}$ and the plasma potential in front of it varies with each electrode. Anticipating our discussion in the next section, we see that this will lead to non-uniform currents to the electrodes.

Irrespective of the biasing case, the plasma potential profiles measured on ports $\#20$ and $\#34$ have almost the same shape, at least in the region in front of the electrodes. In fact one verifies that they can be simply deduced from one another via a uniform shift up (going from $\#34$ to $\#20$) or down (going from $\#20$ to $\#34$). As mentioned above, this shift is consistent with a radially uniform temperature drop between the two axial locations. Other than for this temperature effect, the data in figure~\ref{Fig:t1_FP_6base} shows that the imprint of the bias on the plasma potential profile is recovered at least $6$~m away from the multi-disk electrodes.

\begin{figure}
\begin{center}
\includegraphics[width=13.5cm]{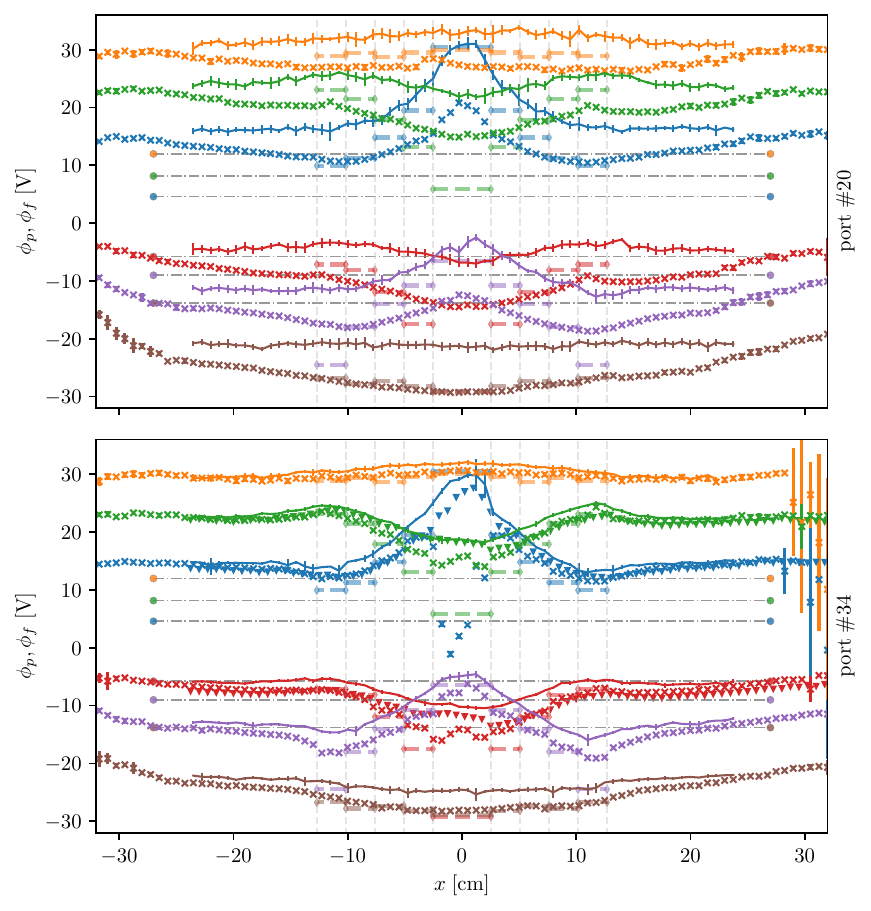}
\caption{Plasma (solid lines) and floating (cross markers) potential profiles on two ports ($\#20$ top, $\#34$ bottom) at $t=15.6$~ms (\emph{i.~e.} in the afterglow during biasing) for the six biasing scenarios listed in table~\ref{Tab:cases}. The vertical dashed lines represent the electrodes position. The thicker and lighter colored horizontal lines represent the electrodes potential $\phi_{i}$ at that time. The horizontal grey dashed lines with coloured circles symbols represent the anode potential $\phi_{a}$. The triangle symbol in the bottom panel correspond to plasma potential measurement from emissive probe on port $\#33$ (when available). }
\label{Fig:t3_FP_6base}
\end{center}
\end{figure}

Finally, we verify in figure~\ref{Fig:t3_FP_6base} that the strong influence of the biased electrodes on the plasma potential, as well as the general trends identified above, persist in the afterglow after the main discharge is shut off. In the afterglow, however, we find that the plasma potential profile is much closer to the floating potential profile. This behaviour is consistent with the rapid drop in $T_{e}$ observed in the afterglow.

\subsection{Density and temperature profiles under active biasing}

\begin{figure}
\begin{center}
\includegraphics[width=13.5cm]{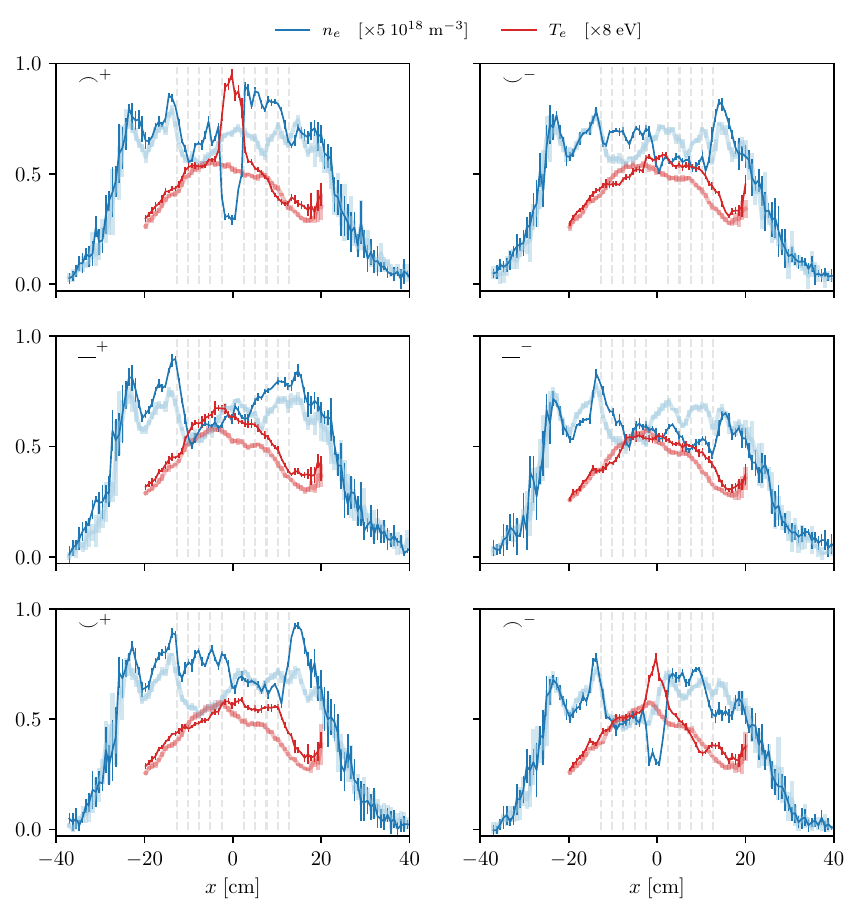}
\caption{Density (blue) and temperature (red) radial profiles on port $\#34$ before active biasing ($t=11.1$~ms, thicker lighter color lines)  and during active biasing in the main discharge ($t=12.6$~ms, thinner solid lines) for the six biasing scenarios listed in table~\ref{Tab:cases} (six different panels). The vertical dashed lines represent the electrodes position. Density and temperature are normalised to $5~10^{18}$~m$^{-3}$ and $8$~eV respectively. }
\label{Fig:DepletionHeating}
\end{center}
\end{figure}

Looking now at the effect of biasing on other plasma parameters, figure~\ref{Fig:DepletionHeating} plots the radial profiles of plasma density $n_{e}$ and of electron temperature $T_{e}$ during active biasing in the main discharge ($t=12.6$~ms), just in front of the electrodes (port $\#34$). 

Starting with the density, one observes for non-uniform biasing scenarios a decrease in density in the region in front of the most positive electrode. This behaviour is most noticeable for the negative gradient biasing scenarios ($\frown^{-}$, bottom-right and $\frown^{+}$, top-left panels in figure~\ref{Fig:DepletionHeating}), for which a clear density drop of $30-50\%$ is observed in front of the central electrode $\mathcal{E}_{1}$. This depletion is associated, as we will show, to an electron current collected on the electrode with the highest potential since, as seen in figure~\ref{Fig:t1_FP_6base}, the sheath in front of this electrode is the smallest. Looking more closely, this localised decrease in density appears to be accompanied by a density increase in the adjacent region. This is notably visible for $x>0$ and the positive gradient scenarios ($\smile^{-}$, top-right and $\smile^{+}$, bottom-left panels in figure~\ref{Fig:DepletionHeating}), for which the density appears to drop when actively biasing for $r\leq12.5$~cm, but to rise for $r>12.5$~cm (\emph{i.~e.} past the outer edge of the outermost electrode $\mathcal{E}_{5}$).

In this same figure we see that this local depletion is accompanied by an increase in the local electron temperature $T_{e}$. This is particularly clear for the negative gradient biasing scenarios ($\frown^{-}$, bottom-right and $\frown^{+}$, top-left panels in figure~\ref{Fig:DepletionHeating}), for which a significant increase in $T_{e}$ is seen in front of the central electrode $\mathcal{E}_{1}$. We note that since the density is here deduced directly from $I_{sat}$, and thus does not account for the effect of a change in $T_{e}$ through the ion sound speed, this local heating will lead to an even stronger density drop.

Exploring how these effects vary axially, that is with the distance from the multi-disk electrode, we found (not shown here) that both the depletion and heating were clearly recovered in the density and temperature profiles measured on port $\#20$, that is about $6$~m away from the electrodes. Finally, these features are observed to persist in the afterglow. In fact the depletion is even more pronounced in the afterglow, with the plasma density then approaching zero. This is because there is in this case no longer an electron beam to act as a plasma source to replenish the depleted flux tube.

\section{Model for the plasma potential response to biasing}
\label{Sec:Model}

Having highlighted in the previous section how different biasing profiles imposed on the multi-disk electrode lead to different plasma potential profiles and anode potential, we now try to shed light on the dynamics at play. We begin by discussing how the different plasma potential profiles observed for the different biasing scenarios can be understood in terms of the axial current drawn at the electrodes. We then show how the current balance at the anode forces the anode potential to respond to the plasma potential created by the multi-disk electrodes. Throughout this section, we assume that the plasma column is axisymmetric with radial profiles consistent with the profiles along $x$ presented in the previous section.

\subsection{Voltage drop along field lines}

The first element to analyse the evolution of the plasma potential with biasing is to appreciate that although it is small, the non-zero axial plasma resistivity can be the source of significant (with respect to the applied bias) axial voltage drops.  

To see this recall that, for the low operating pressure used on LAPD, the dominant contribution to parallel resistivity is electron-ion collisions~\citep{Poulos2019}. The parallel resistivity is thus the parallel Spitzer resistivity, 
\begin{equation}
\eta_{ei} = 1/\sigma_{\parallel}^{ei} =\frac{1}{1.96}\frac{m_{e}\nu_{ei}}{n_{e}e^{2}}
\label{Eq:Spitzer_res}
\end{equation}
with 
\begin{equation}
\nu_{ei} = \frac{n_{e}Z^{2}e^{4}\Lambda_{c}}{6\sqrt{2}\pi^{3/2}\epsilon_{0}^{2}\sqrt{m_{e}}(k_{B}T_{e})^{3/2}}.
\end{equation}
the Coulomb collision frequency and $\Lambda_{c}$ the Coulomb logarithm. For the typical plasma parameters given in table~\ref{Tab:baseline} one gets $\eta_{ei}\sim7~10^{-5}~\Upomega.$m. From there one finds that the voltage drop along field lines over a distance $s$ for a given parallel current $j_{\parallel}$ is 
\begin{equation}
\Delta_{\parallel}\phi_{p}(r) = \int_{0}^{s} \eta_{ei} j_{\parallel}(r,z)dz.
\label{Eq:axial_drop}
\end{equation}
Using as an estimate the characteristic ion saturation current density $j_{is}^{\diamond}=8.7$~kA.m$^{-2}$ defined in Eq.~\eqref{Eq:j_is_diamond} this leads to $\eta_{ei} j_{is}\sim0.6~$V.m$^{-1}$. Integrating over the $11$~m separating the anode from the multi-disk electrode, this would lead to $\Delta_{\parallel}\phi_{p}\sim7$~V, which is indeed significant compared to the $\pm30$~V of applied bias.

\subsection{Current driven plasma potential radial profile}

\begin{figure}
\begin{center}
\includegraphics[width=13cm]{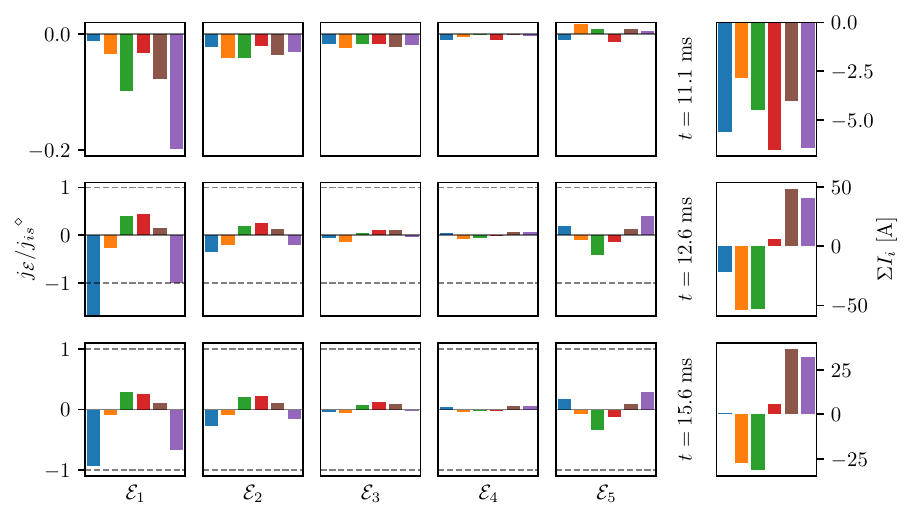}
\caption{Current density $j_{{\mathcal{E}}}$ on each electrode $\mathcal{E}_{i}$ for $i\in\llbracket1,5\rrbracket$ ($5$ leftmost columns) and total current $\Sigma I_{i}$ drawn by the multi-disk electrode (rightmost column) at the three different instants of interest (rows) and for all biasing cases (color code) shown in table~\ref{Tab:cases}. The current density is here normalized by the characteristic current density ${j_{is}}^{\diamond}$ (see Eq.~\eqref{Eq:j_is_diamond}), and obtained by assuming a uniform current distribution on each electrode. The horizontal dashed gray lines highlights $j_{{\mathcal{E}}}=\pm{j_{is}}^{\diamond}$. }
\label{Fig:Current_electrodes}
\end{center}
\end{figure}

To confirm this finding and the importance of a voltage drop along field lines in the quasi-neutral plasma, one still needs to know how large the actual parallel current $j_{\parallel}$ is compared to the characteristic ion saturation current density $j_{is}^{\diamond}$. To do so, we assume current conservation along magnetic field lines to use here as a proxy the current density, $j_{\mathcal{E}_{i}}$ which is deduced from the current $I_{i}=j_{\mathcal{E}_{i}}\mathcal{A}_{i}$ collected on each electrode $\mathcal{E}_{i}$ of surface $\mathcal{A}_{i}$. This dataset is shown in figure~\ref{Fig:Current_electrodes}. Beyond a substantial overall increase in current drawn when bias is turned on, one verifies that current densities $|j_{\mathcal{E}_{i}}|$ exceeding $j_{is}$ are collected on the innermost electrode $\mathcal{E}_{1}$ when this electrode is the most positive (negative potential gradients on the electrodes, blue and purple). This confirms that the axial currents that result from active biasing are indeed the source of voltage drops along field lines of more than $10$~volts.

\begin{figure}
\begin{center}
\subfigure[$t=12.6$~ms]{\includegraphics[height=6.75cm]{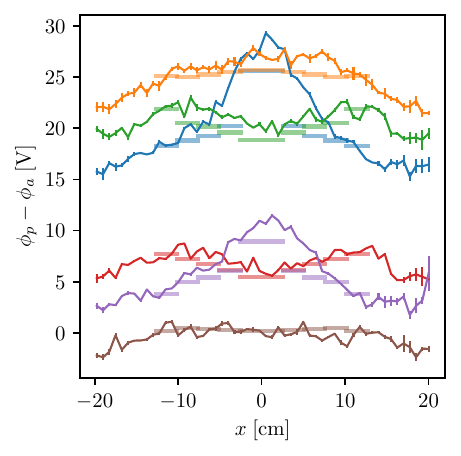}}\subfigure[$t=15.6$~ms]{\includegraphics[height=6.75cm]{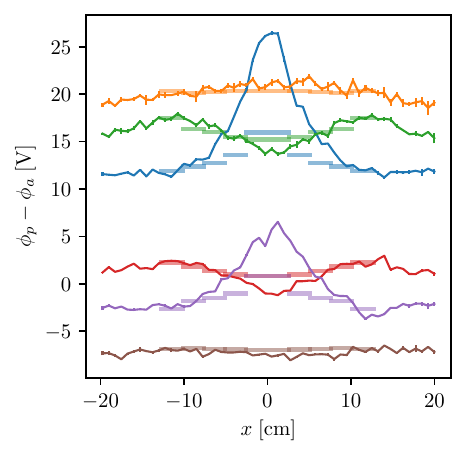}}
\caption{Radial profile of the plasma potential $\phi_{p}(r,z_{20})$ minus the anode potential $\phi_{a}$ measured at $t=12.6$ (main discharge, left) and $t=15.6$~ms (afterglow, right) on port $\#20$ (solid lines with error bars) compared with the profiles constructed from an offset $\varphi$ and the voltage $\eta_{ei}j_{\mathcal{E}_{i}}l$ (thick lines, see Eq.~\eqref{Eq:fit}), with $j_{\mathcal{E}_{i}}$ the current density measured on electrode $\mathcal{E}_{i}$ at this instant, $l=z_{a}-z_{20}=6$~m for the distance between the anode and port $\#20$, and $\eta_{ei}=7~10^{-5}~\Upomega.$m which corresponds to the baseline parameters given in table~\ref{Tab:baseline}. }
\label{Fig:Fit}
\end{center}
\end{figure}

Exploring this idea further, we can examine how the radial dependence of $j_{{\mathcal{E}}}$ can possibly explain the observed radial profile of the plasma potential $\phi_{p}(r)$. To do so, we simply construct the radially dependent axial voltage drop $\Delta_{\parallel}\phi_{p}(r)$ computed from a constant and uniform resistivity $\eta_{ei}$ and the time-dependent and electrode dependent $j_{{\mathcal{E}_{i}}}$. Comparing these simple results with the actual plasma potential profiles $\phi_{p}(r)$ as done in figure~\ref{Fig:Fit}, we see that this simple model reproduces very well the radial variations of the plasma potential. Specifically, figure~\ref{Fig:Fit} shows that 
\begin{equation}
\phi_{p}(r,z_{20}) \approx \phi_{a} + \varphi + \eta_{ei}j_{\mathcal{E}_{i}}(z_{a}-z_{20}),
\label{Eq:fit}
\end{equation}
where $\varphi$ is an offset and $\phi_{a}$ is the anode potential. Although, as summarized in table~\ref{Table:offset} $\phi_{a}$ and $\varphi$ are both biasing scenario dependent, they are importantly radially uniform. The remarkable fit observed in figure~\ref{Fig:Fit} thus supports further the idea that the plasma potential profile is driven by the axial currents $j_{\mathcal{E}_{i}}$ collected on biased electrodes. Radial potential shaping in the plasma is thus achieved by drawing different axial currents on the different electrodes. On the other hand the reason for this offset remains at this point unclear. To answer this question, we need to look at the anode-cathode system and how its dynamics is coupled to that of the plasma, as we will do next.

\begin{table}
\begin{center}
\begin{tabular}{c | c c | c c}
Biasing scenario  & \multicolumn{2}{c|}{Offset $\varphi$~[V] } & \multicolumn{2}{c}{Anode potential $\phi_{a}$~[V]}\\
& $12.5$~ms & $15.5$~ms & $12.5$~ms & $15.5$~ms\\
$\frown^{+}$\quad\textcolor{tabblue}{\rule{.4cm}{0.1cm}} & $18$ & $12$ & $5$  & $4$\\
\rule{0.3cm}{0.015cm}$^{+}$\quad\textcolor{taborange}{\rule{.4cm}{0.1cm}} & $25$ & $20$ & $11$  & $12$\\
$\smile^{+}$\quad\textcolor{tabgreen}{\rule{.4cm}{0.1cm}} & $22$ & $17$ & $8$  & $8$\\
$\smile^{-}$\quad\textcolor{tabred}{\rule{.4cm}{0.1cm}} & $8$ & $2$ & $-5$  & $-6$\\
\rule{0.3cm}{0.015cm}$^{-}$\quad\textcolor{tabbrown}{\rule{.4cm}{0.1cm}} & $0$ & $-7$  & $-11$ & $-14$\\
$\frown^{-}$\quad\textcolor{tabpurple}{\rule{.4cm}{0.1cm}} & $4$ & $-3$ & $-8$  & $-9$\\
\end{tabular}
\caption{Offset $\varphi$ to be added to the anode potential $\phi_{a}$ and the axial voltage drop $\eta_{ei}j_{\parallel}l$ to reproduce the observed radial plasma potential profile $\phi_{p}(r,-l)$ on port $\#20$. For comparison $\varphi=11\pm1$~V for all scenarios at $t=11.6$~ms, that is prior to active biasing.}
\label{Table:offset}
\end{center}
\end{table}

Note finally that we could have captured the radial and biasing scenario dependencies of $\eta_{ei}$ via the measured profiles of $T_{e}$. Our choice to simply use here a constant value $\eta_{ei}=7~10^{-5}~\Upomega.$m is motivated by our goal in this study to highlight global trends rather than to do a quantitative analysis.

\subsection{Current balance at the anode}

We now would like to explain the offset $\varphi$ needed to reproduce the plasma potential in Eq.~\eqref{Eq:fit}, and because this offset is added to the anode potential $\phi_{a}$ we must also consider the anode dynamics. 

A key element to understanding the response of the plasma-anode system to biasing is the overall current balance in the machine. During the main discharge, an electron current $\iota I_{d}$, with $\iota$ the anode transparency, is injected through the mesh anode into the main chamber. This is equivalent to the current density $j_{inj}$ defined in Eq.~\eqref{Eq:j_inj}. Although the anode-cathode system is floating with respect to the machine ground~\citep{Qian2023}, this injected current $-\iota I_{d}$ must be balanced by an equal excess of electron current returning to the anode from the plasma in the main chamber. This in turn sets constraints on the voltage drop one can have between the anode potential $\phi_{a}$ and the plasma potential in front of the anode $\phi_{p}(r,z_{a})$. Indeed, assuming an ion sheath, the current density collected on the anode is~\citep{Poulos2019,Trotabas2022} 
\begin{equation}
\label{Eq:current_ion_sheath}
j_{a}(r) = j_{is}\left(1-\exp\left[\Lambda+\frac{\phi_{a}-\phi_{p}(r,z_{a})}{T_{e}}\right]\right)
\end{equation}
with 
\begin{equation}
\Lambda=\ln\left[\sqrt{\frac{m_{i}}{2\pi m_{e}}}\right]
\end{equation}
a parameter that is about $3.5$ in helium. The anode potential $\phi_{a}$ and the plasma potential in front of the anode $\phi_{p}(r,z_{a})$ must thus adjust so that overall
\begin{equation}
\label{Eq:balance_anode}
j_{inj}\pi\rho_{B}r_{c}^{2} = \int_{0}^{{\sqrt{\rho_{b}}r_{c}}}2\pi r dr j_{a}(r).
\end{equation}
Note importantly that this condition is global as opposed to local, so that there can still be current along magnetic field lines, \emph{i.~e} $j_{a}(r) \neq j_{inj}$. 

We now see the origin of the offset $\varphi$ in Eq.~\eqref{Eq:fit}. It is the height of the anode sheath required to ensure the needed electron return current $-\iota I_{d}$ to the anode.

\subsection{Effect of biasing on the anode sheath height}

To see why the anode sheath and thus the offset $\varphi$ depend on the biasing scenario, we need to consider the total current balance. 

In the absence of parallel current, we can safely disregard the dynamics at the multi-disk electrode and focus exclusively on the anode.  Assuming further a uniform plasma potential in front of the anode, $j_{a}=j_{inj}$ and Eq.~\eqref{Eq:current_ion_sheath} then gives 
\begin{equation}
\varphi \approx \left[\Lambda-\log\left(1-\frac{j_{inj}}{j_{is}}\right)\right]T_{e},
\end{equation}
where we recall from Eq.~\eqref{Eq:j_inj} that $j_{inj}<0$. For comparable current densities $j_{inj}$ and $j_{is}^{\diamond}$ this would lead to a voltage drop of $2.5-3~T_{e}$.

This regime of negligible parallel current happens to be relatively representative of the regime before active biasing ($t\leq12$~ms). As shown in the first panel in figure~\ref{Fig:Current_electrodes}, the current density on the electrodes is indeed much smaller than to the ion saturation current density $j_{is}^{\diamond}$. From Eq.~\eqref{Eq:axial_drop} this implies a negligible voltage drop along field lines, so that we can in first approximation consider $\phi_{p}(r,z_{a})\approx\phi_{p}(r,z_{20})$. Now, looking back at the left panel in figure~\ref{Fig:t0_FP_6base}, we find that the plasma potential on port $\#20$ is about $10$ to $15$~V higher than the anode potential $\phi_{a}$. This voltage drop is relatively consistent with the $2.5-3~T_{e}$ predicted above, given the temperature $6\leq T_{e}\leq8$~eV measured on the port $\#20$. It is also very consistent with the offset $\varphi=11\pm1$~V computed at this instant for all biasing scenarios, supporting the interpretation of $\varphi$ in terms of the anode sheath height.

In contrast, we have seen in the middle panel of figure~\ref{Fig:Current_electrodes} that current densities comparable or even larger in amplitude than $j_{is}^{\diamond}$ are collected on the multi-disk electrode during active biasing. Beyond leading, as discussed earlier, to a significant axial voltage drop along the field line in the plasma, this collected current also modifies the total current balance. A negative current drawn on the multi-disk electrode must be compensated by a decrease in the electron current collected at the anode, which is achieved by increasing the sheath height. Going back to table~\ref{Table:offset} and the last column of figure~\ref{Fig:Current_electrodes}, we verify that the offset indeed is larger for positive biases (blue, orange and green), and that the larger the amplitude of the total current drawn $\Sigma I_{i}<0$, the larger the offset. Conversely, a positive current drawn on the multi-disk electrode, which for an ion sheath implies a reduced electron current, must be compensated by an increase of the electron current collected at the anode, which is achieved by decreasing the sheath height. Going once again back to table~\ref{Table:offset} and the last column of figure~\ref{Fig:Current_electrodes}, we verify this time that the offset is indeed smaller for negative biases (red, brown and purple), and that the larger the amplitude of the total current drawn $\Sigma I_{i}>0$, the smaller the offset. Finally, the offset $\varphi$ pre-biasing, for which the potential $\phi_{i}$ on the electrodes is as shown in figure~\ref{Fig:t0_FP_6base} only a few volts below ground, seats as expected in between those of positive and those of negative biases.

Finally, we note in table~\ref{Table:offset} that the offset changes, and more precisely decreases slightly, when moving into the afterglow. This behaviour was to be expected as there is in this case no longer current injection at the anode, and thus no longer the need to collect this current back to the anode from the plasma.

\subsection{Discussion}

In summary, the analysis proposed here of the experimental results obtained in this campaign indicate that the plasma potential profile is primarily controlled via the boundary condition imposed on a given field line by the multi-disk electrode. This suggests that the dynamics is essentially axial (\emph{i.~e} parallel to the background magnetic field), which is consistent with earlier results on LAPD by~\cite{Jin2019}, and with the analysis of these results proposed by~\cite{Poulos2019} in the limit that $\tau\ll1$ (see Eq.~\eqref{Eq:tau}). This also suggests that more pronounced profiles, \emph{e.~g.} stronger radial electric field, could be produced by drawing appropriate currents on the biased end-electrodes. 

A few comments are however called for at this point. \cite{Jin2019} used hot electrodes producing parallel current densities a few times the ion saturation current density (in amplitude), which is significantly larger than those measured for cold electrodes in this study. This weaker current could have suggested a larger voltage drop across the sheath~\citep{Poulos2019,Trotabas2022}, contrary to what was observed in the results reported here. The reason for this finding is twofold. First, in biasing with respect to the chamber wall as opposed to the anode, the anode was here free to self-adjust, limiting in turn the amplitude of the voltage drop across the sheath formed in front of the multi-disk electrode for negative biases. Second, although they remain smaller than those produced by hot electrodes, the parallel currents generated here for positive biases were large enough to support axial voltage drops along field lines that are comparable to the applied bias, limiting again the voltage drop across the sheath.

\section{Biasing driven plasma rotation}
\label{Sec:Rotation}

Having established in the previous sections that differentially biasing the multi-disk electrodes with respect to the machine ground has a clear effect on the radial profile of plasma potential $\phi_{p}(r)$ in the machine, we can now infer how these different equilibria lead to different rotation profiles. 

\subsection{Cross-field rotation from end-electrodes biasing}

In a linear machine with $\mathbf{B}=B_{0}\mathbf{\hat{e}}_{z}$ radial gradients of pressure $p(r)$ and plasma potential $\phi_{p}(r)$ are the source of different azimuthal drifts that can both contribute to plasma rotation. One is the diamagnetic drift $\bm{\Omega}_{D}=\Omega_{D}\mathbf{\hat{e}}_{z}$ where
\begin{equation}
\Omega_{D} =-\frac{|\bm{\nabla} p\times\mathbf{B_{0}}|}{rqnB_{0}^{2}} = \frac{1}{rqnB_{0}}\frac{\partial p(r)}{\partial r}
\end{equation}
with $q$ the signed particle charge. The other is the $\mathbf{E}\times\mathbf{B}$ drift, which leads to a cross-field angular plasma frequency $\bm{\Omega}_{E\times B}=\Omega_{E\times B}\mathbf{\hat{e}}_{z}$ with
\begin{equation}
\Omega_{E\times B} =\frac{1}{rB_{0}}\frac{\partial \phi_{p}(r)}{\partial r}. 
\end{equation}
A positive potential radial gradient thus leads to $\Omega_{E\times B}>0$ and a cross-field drift in the direction of the electron diamagnetic drift (assuming $\partial p/\partial r<0$) while a negative potential radial gradient leads to $\Omega_{E\times B}<0$ and a cross-field drift in the ion diamagnetic drift. 

Comparing these two contributions, one finds 
\begin{equation}
\frac{\Omega_{E\times B}}{\Omega_{D}}=\frac{e}{kT}\frac{\partial \phi_{p}}{\partial r}\bigg/\left[\frac{1}{T}\frac{\partial T}{\partial r}+\frac{1}{n}\frac{\partial n}{\partial r}\right]
\end{equation}
where $\phi_{p}, n$ and $T$ are all functions of $r$. This shows that the $\mathbf{E}\times\mathbf{B}$ contribution dominates over the diamagnetic contribution if the drop in plasma potential over the characteristic gradient lengths for the density and the temperature is equal or larger than a few $T_{e}$. Because we observe radial voltage drops as large as $2$ to $3~T_{e}$ across the radius of the multi-disk electrode while the density and temperature vary by at most $30\%$ in this same region, this assumption appears justified here, at least in the central region. 

\begin{figure}
\begin{center}
\subfigure[$t=12.6$~ms]{\includegraphics[width=6.75cm]{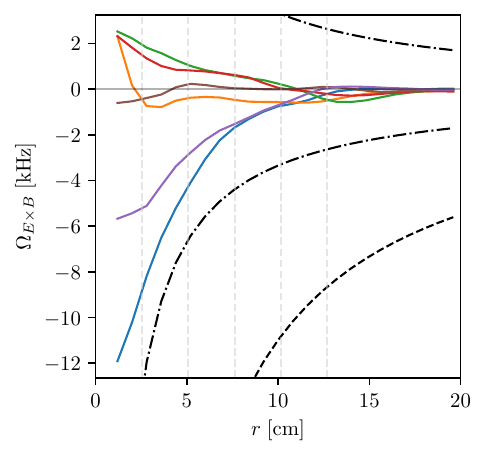}\label{Fig:Omega_12.5}}\subfigure[$t=15.6$~ms]{\includegraphics[width=6.75cm]{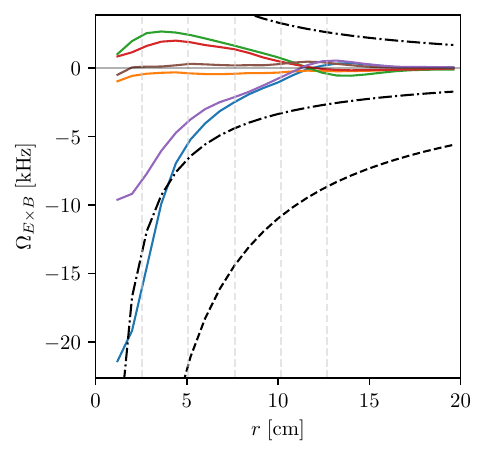}\label{Fig:Omega_15.5}}
\caption{Angular plasma frequency on port $\#34$, that is about $30$~cm away from the multi-disk electrode, at $t=12.6$~ms~\subref{Fig:Omega_12.5} and  $t=15.6$~ms~\subref{Fig:Omega_15.5} for the six biasing scenarios (colour code) listed in table~\ref{Tab:cases}. Rotation is computed from the plasma potential profiles assuming pure $\mathbf{E}\times\mathbf{B}$ rotation. The dashed and dash-dotted black curves represent sonic rotation for $T_{i}=1$ and $0.1$~eV, respectively. The vertical grey dashed lines show the electrodes position.}
\label{Fig:12.5_Omega}
\end{center}
\end{figure}

Under this simplifying assumption, it is possible to infer rotation from the plasma potential profiles measured for the different biasing scenarios. To do so we fit the plasma potential profiles using cubic smoothing splines, then fold back the negative $x$ over the positive $x$ to obtain an average radial profile $\phi_{p}(r) = [\phi_{p}(x=-r<0)+\phi_{p}(x=r>0)]/2$, which is finally differentiated with respect to $r$ to get minus the radial electric field. The results obtained for active biasing in the main discharge ($t=12.6~$ms) and in the early afterglow ($t=15.6$~ms) are shown in figure~\ref{Fig:12.5_Omega}. 

As anticipated from the plasma potential profiles, we find the largest rotation in the core for the two negative gradient biasing scenarios (blue and purple) for which electrons are drawn by the electrodes on the axis. Rotation is counterclockwise, that is in the ion diamagnetic direction. Although rotation likely remains subsonic, we see on this same figure that the azimuthal drift locally approaches what would be the sonic speed for an ion temperature of a fraction of an eV. We also verify that the non-monotonic plasma potential profiles measured for the two positive gradient biasing scenarios (green and red) lead to a rotation reversal near the edge of the outermost electrode. Rotation is in this case clockwise in the core, and counterclockwise in the outer region. Finally, uniform biasing profiles (orange and brown), which we have seen lead to comparatively flat plasma potential profiles, consistently yield weak rotation.

Note however again that this analysis assumed $|\Omega_{E\times B}|\gg|\Omega_{D}|$. Clearly this is not valid for non-monotonic profiles near the point of reversal, where by definition $\Omega_{E\times B}\sim 0$. This ordering is also questionable for flat potential profiles. In short, the examination of the potential profiles points to a strong plasma rotation driven by biasing for the negative gradient biasing scenarios, but a more in-depth analysis of the rotating equilibrium is called for other biasing scenarios.

\subsection{Discussion}

Having underlined the possibility to produce rapidly rotating plasmas via end-electrodes biasing, we now briefly discuss this result in light of related experiments previously conducted in LAPD. Of particular interest here is the work by~\cite{Maggs2007} and~\cite{Carter2009} in which a section of the chamber wall was biased positively with respect to the cathode to drive rotation. 

An interesting observation is that although both biasing schemes may lead (did lead in the case of ~\cite{Maggs2007}) to rotation, they do so in very different ways. In the experiment discussed here, we surmised that rotation results from axial currents drawn by end electrodes, which then support a radial electric field. In this case, we have seen that rapid rotation appears more easily achievable in the counterclockwise direction ($E_{r}>0$). On the other hand, in~\cite{Maggs2007} it is an inward ion radial current driven by a radial voltage difference that led to a clockwise ($E_{r}<0$) plasma rotation. Biasing the chamber negatively compared to the cathode, which would have led to counterclockwise rotation, was tried but shown to result in arcing. These basic differences are then expected to lead to distinct behaviours, which future studies may be able to examine. 

First, the axial voltage drop model proposed here to explain end-electrodes biasing results would lead to a radial electric field that varies along $z$, decreasing in amplitude from the multi-disk electrode to the anode. The corresponding axially dependent crossed-field rotation would in turn translate into a twist of the plasma. On the contrary, the edge biasing, if done over the entire length of the plasma column, would in principle lead to an axially uniform rotation. Second, with end-biasing, the axial voltage drop and thus the radial electric field amplitude scale with the parallel Spitzer resistivity Eq.~\eqref{Eq:Spitzer_res}. Meanwhile, in edge biasing experiments ions are transported radially due to ion-neutral collisions, and the radial electric field is hence inversely proportional to the ion-driven Pedersen conductivity
\begin{equation}
\sigma_{p} = \frac{n_{e}m_{i}\nu_{in}}{{B_{0}}^2}
\end{equation}
with $\nu_{in}$ the ion-neutral collision frequency. Comparing these two scaling, one expects completely different dependencies on plasma parameters. 

Lastly, an important finding from~\cite{Maggs2007} was that rotation driven by biasing led to a modification of transport properties, and more specifically to a reduction of cross-field transport from Bohm to classical rates, which translated into a steepening of the plasma density radial profiles. Although this is not the focus of our study, we note here in contrast looking back at figure~\ref{Fig:DepletionHeating} that there does not appear to be a noticeable change in density profile in the end-electrodes biasing results reported here.

\section{Conclusion}
\label{Sec:Conclusion}

An experimental campaign was carried out on the Large Plasma Device (LAPD) at the University of California, Los Angeles to study how biased end-electrodes can be used to control the radial potential profiles in the plasma bulk in the machine. It uses a set of five independently biased disk electrodes to control and impose boundary conditions for the potential about $11$~m from the anode. Typical operating conditions in the main discharge were $n_{e}=5~10^{18}$~m$^{-3}$ and $T_{e}=6-8$~eV.

Under these conditions, we observed that biasing has a very noticeable effect on the plasma potential in the machine, at least up to $6$~m from the electrodes, with shifts in plasma potential that are comparable to the $\pm30$~V amplitude bias imposed on the electrodes. Examining the plasma response to non-uniform (though monotonic) biasing profiles, it was found that concave-down biasing profiles lead to the largest radial drops in plasma potential, no matter whether the applied bias is positive or negative. On the other hand, concave-up biasing profiles were shown to lead to non-monotonic profiles and less pronounced plasma potential radial gradients.

Analysing these results, it was shown that the radial profile of plasma potential in the machine is governed by the current drawn by each of the concentric electrodes and the axial resistance associated with Spitzer conductivity. The largest radial variations observed for concave-down biasing profiles are notably driven by the larger electron current drawn by the on-axis electrode in this case, suggesting that steeper profiles may be achievable if drawing suitable parallel currents. It was also found that the nature of the plasma source on LAPD, namely the injection of energetic electrons through the anode, further imposes constraints between the plasma potential and the anode potential, which in turn forces the anode potential to track variations in plasma potential.

Finally, we computed the cross-field rotation expected from the different plasma potential profiles resulting from the various biasing scenarios. This analysis suggests that concave-down biasing profiles on end-electrodes may be used to produce rotating plasmas that locally approach sonic rotation, complementing other biasing-driven rotation techniques that had been previously demonstrated on LAPD. On the other hand, in producing non-monotonic potential profiles, concave-up biasing profiles would lead to flow reversal.  In this latter case, though, one expects diamagnetic contributions to modify this picture.

\section*{Acknowledgments}
The authors would like to thank Patrick Pribyl for his guidance on pulser biasing and emissive probes, Vadim Murinov for his help developing a script to extract electron temperature data from Langmuir scans, as well as Zoltan Lucky, Marvin Drandell, and Tai Ly for their technical support and assistance. The authors would also like to thank Troy Carter and Stephen Vincena for constructive discussions of the experimental results, and Stewart J. Zweben for his help designing this experiment.

\section*{Funding}
This research was performed at the Basic Plasma Science Facility at UCLA, which is funded by the United States Department of Energy and the National Science Foundation. RG gratefully acknowledges financial support for travel expenses from the International Action Program from CNRS INSIS.

\section*{Declaration of interests}
The authors report no conflict of interest.



\end{document}